# Route Planning Made Easy – An Automated System for Sri Lanka

M F M Firdhous[1], D L Basnayake[2], K H L Kodithuwakku[3], N K Haththalla[4], N W Charlin[5], P M R I K Bandara[6]

Faculty of Information Technology
University of Moratuwa
Sri Lanka

[1]firdhous@uom.lk, {[2]dlbs.lk, [3]harshanikhl, [4]nimali.devi, [5]nwcharlin, [6]isuri86}@gmail.com

*Abstract – Commercial cities like Colombo constantly face to problem of traffic congestion due to the large number of people visiting the city for various reasons. Also these cities have a large number of roads with many roads connecting any two selected locations. Finding the best path between two locations in Colombo city is not a trivial task, due to the complexity of the road network and other reasons such as heavy traffic, changes to the road networks such as road closures and one-ways. This paper presents the results of a study carried out to understand this problem and development of a system to plan the travel way ahead of the planned day or time of the journey. This system can compute the best route from between two locations taking multiple factors such as traffic conditions, road closures or one-way declarations etc., into account. This system also has the capability to compute the best route between any two locations on a future date based on the road conditions on that date. The system comprises three main modules and two user interfaces one for normal users and the other for administrators. The Administrative interface can only be accessed via web browser running on a computer, while the other interface can be accessed either via a web browser or a GPRS enabled mobile phone. The system is powered mainly by the Geographic Information System (GIS) technology and the other supporting technologies used are database management system, ASP.Net technology and the GPRS technology. Finally the developed system was evaluated for its functionality and user friendliness using a user survey. The results of the survey are also presented in this paper.*

I. INTRODUCTION

Commercial cities like Colombo have a large floating population in addition to the population who live there permanently. People come to Colombo for various reasons such as official, business and personal work. Also, people living within the Colombo city move from one place to another for similar purposes. Also, Colombo is the main port of Sri Lanka. Hence almost all the import into the country and export out of the country need to go through Colombo. In addition to the import and export, local produce such as vegetable, fish and other consumer goods arrive at Colombo daily from the provinces. All these movement of human and goods naturally make Colombo the busiest city in the country. Whatever the reason, people and vehicle move from one place to another, they like to reach the intended destination as soon as possible with minimum hassle. But, with the present conditions within the city, most of them will not succeed due to variety of reasons, such as vast amount of roads within the city, traffic congestion, regular and unannounced closure of roads, accidents, etc., (Karunaratne, 2006). Hence a route planning and advisory system would be of immense help to the people who travel to and within Colombo. A route advising system that can determine the best route from a starting point another point within the city along with proper driving instructions will not only help people to select the best route, this would also increase the productivity of users as they would be able to complete their work within the shortest possible time while saving the country a lot of foreign exchange in terms of reduced fuel consumption in vehicles. Considering these facts, it was decided to build a system to address the above issue using common and easy to use access methods such as the World Wide Web and the mobile phone which are very popular among people today.

The authors of this paper propose a system solute to design and develop a Route Planning and Advising System that will help the users to select the best route between two points within the Colombo city. The system has been built using the three-tier architecture comprising of backend database, business logic and the presentation layer. The backend database stores route and other relevant information, business logic module carries out the necessary calculations while the presentation layer would present the result to the user in the user selected format. The presentation layer has been developed on two independent modules namely the web module that provides the user interface on any regular web browser and the mobile module that supports the mobile user interface.

The main reason for developing the presentation layer is the popularity of both technologies in Sri Lanka. Sri Lanka has a large computer literate population compared to other countries in the region and the penetration of the mobile phones in Sri Lanka in one of the highest in the world compared to its population. Figure 1 shows the growth of Internet and email users over the years from its introduction in the mid 1990s.

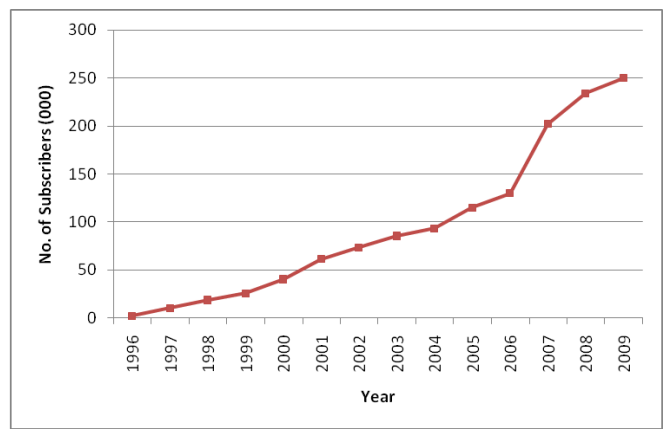

Figure 1: Number of Internet and Email Users over the Years
Data Source: Telecommunications Regulatory Commission of Sri Lanka

Figure 2 shows the growth of mobile phone subscribers in Sri Lanka starting from early 1990s. From this figure, it can be seen that the total number of subscribers has reached 14 million in 2009. Presently there are five mobile phone operators offering the latest 3G services covering almost the entire country. Hence, introducing a new application on a web or a mobile will not be a problem as the required infrastructure is already in place.

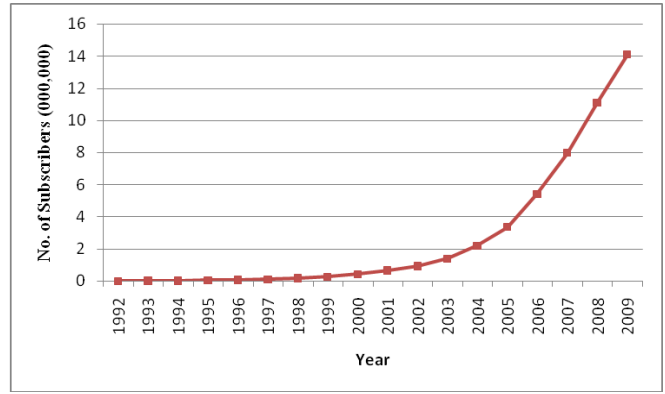

Figure 2: Number of mobile phone users over the Years
Data Source: *Telecommunications Regulatory Commission of Sri Lanka*

The web application will allow the user to select both the starting and end points on a map or using text inputs. Once the user has selected the inputs, the nearest locations to the points selected will be retried from the database and the driving instructions would be presented along with a map. The mobile user is also presented with a map along with spaces to enter the location details similar to the web interface. But the result will be presented only on a map on a mobile phone, whereas the web user would be getting the map as well as detailed directions in the text format. The map information has been processed using the Map Server (GIS Server) running on the Internet Information Services (IIS) (McKenna et al., 2009). C# map scripts are used for querying the Map server while the database is hosted on PostgreSQL with PostGIS. Mobile connector acts as the interface between the mobile phone/desktop client and the GIS server.

The paper has been divided into seven sections where Section I introduces the paper, a detailed discussion on the related work in the area along with their advantages and disadvantages are presented in Section II. Section III presents the overall design of the Route Planning System including technologies used and user management. Section IV discusses the analysis and the design of the system while Section V discusses the implementation of the system. Section VI presents the user feedback on the evaluation of the system for the features provided and user friendliness. Section VII presents the conclusion and recommendations future work.

## II. RELATED WORK

Currently a number of websites host detailed maps of the Sri Lankan cities and directions to famous places for the travellers (SLMAP1, 2009; SLMAP2 2009; SLMAP3, 2009; Watugala 2005). The target audience of these websites are tourists travelling to Sri Lanka. They provide details about the attractions available in Sri Lanka such as sunny beaches, upcountry, historical sites or any other tourist attraction. In terms of driving direction, these websites limit the details to how to reach the places mainly the nearest city from Colombo. Hence, none of this is useful for a person travelling from one place to another within a city like Colombo. The other disadvantage of these sites is that these are static images that lack information on road conditions such as one-way, two-way or totally closed for vehicular traffic. The main advantage of the system discussed in this paper over the other web sites is that any user can find the most appropriate path/direction from one location to the other within the city selected. This is very useful for the local community and as well as tourists, as the system considers all the possible routes between the two locations and provides the most appropriate route for a particular user.

In addition to the local maps described above, there are very popular mapping applications like Yahoo Maps, Google Maps, Bing Maps, MapQuest, and Ovi Maps. These map services also provide driving directions from one place to another (Yahoo, 2009; Google, 2009; Bing 2009; MapQuest 2009; Ovi 2009). Figures 3, 4 and 5 show the map of a part of Colombo city provided by Yahoo Maps, Google Maps and Bing Maps respectively. From these figures, it can be seen that except the Google Map, other two do not have any details on the road network of Colombo. Table 1 compares the features provided by the different mapping services.

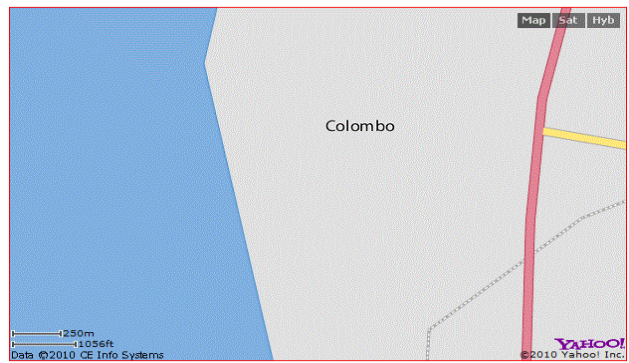

Figure 3: Colombo City Map Provided by Yahoo

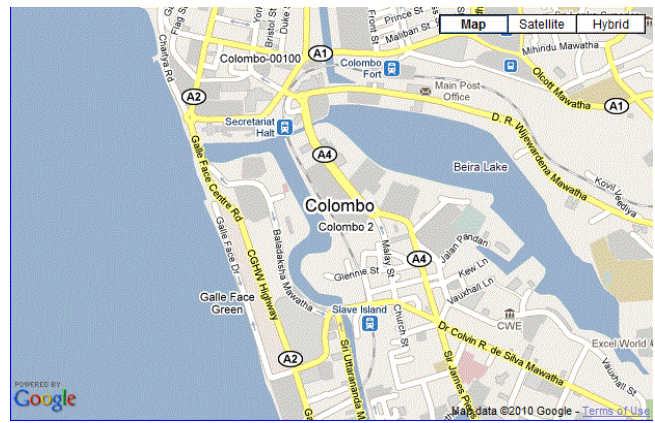

Figure 4: Colombo City Map Provided by Google

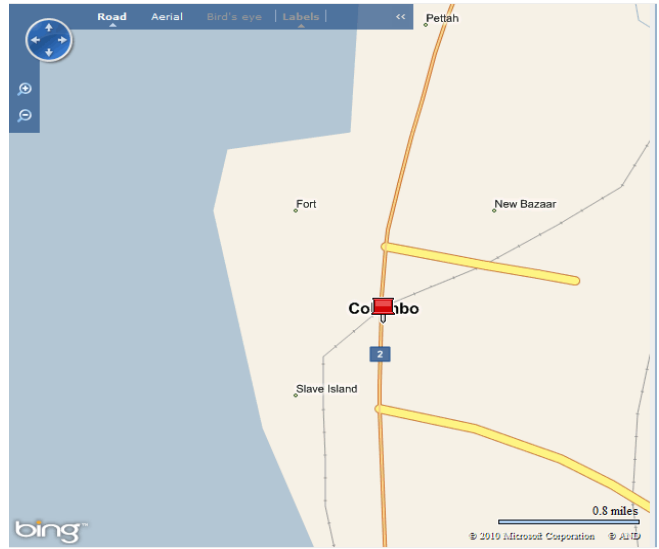

Figure 5: Colombo City Map Provided by Bing

Table 1: Comparison of Features of Different Map Service

| Feature | Yahoo Maps | Google Maps | Bing Maps | MapQuest |
|---|---|---|---|---|
| Map View | Yes | Yes | Yes | Yes |
| Detailed Map (Sri Lanka) | No | Yes | No | No |
| Driving Direction (Sri Lanka) | No | Yes | No | No |
| Reverse Direction (Sri Lanka) | No | Yes | No | No |
| Mobile Option | Yes | Yes | Yes | Yes |
| Dynamic Conditions | No | No | No | No |
| Locally Applicable | No | Somewhat | No | No |
| Time Dependent Results | No | No | No | No |
| Alert Capability | No | No | No | No |

Source: Prepared by Authors

From the comparison of the different map service in Table 1, it can be seen that only Google Map has somewhat local applicability for Sri Lanka. Even though Google

Maps has excellent map and driving direction details, it takes only the permanent conditions of the road such as permanently closed for vehicular traffic or one way etc., Dynamic conditions of the road such as traffic congestion during peak hours such as morning, afternoon or school leaving hours are not taken into account into selecting the best route between the given points. Also, Google does not have the feature of computing the best route for a future time given or alerting system via email or SMS when the road conditions suddenly change due to ad-hoc road closures or accidents. These features are very important for Sri Lankan travellers as road conditions are highly unpredictable and traffic jams a frequent scenario in Sri Lanka.

The objective of this project was to make a system that is suitable for Sri Lanka along with the additional features. The roads in the major cities in Sri Lanka like Colombo are adjusted to meet the requirements of security and traffic conditions. That is, certain roads are made one-way during particular hours of the day while the other times they are two-way roads carrying traffic in both directions. Similarly certain roads are permanently closed for vehicular traffic during particular times or days. Hence, the system developed for Sri Lanka should be capable of taking this additional information into consideration when computing the best route between two given points. Also, the system should help travellers top plan their journeys on a future time and date based on the road conditions on that particular day.

## III. ROUTE PLANNING SYSTEM

*Technologies Used*

The main technologies used to implement the Route Planning System are the GIS and Microsoft ASP.NET. MapServer for Windows has been selected as the GIS software and the other technologies used are postGIS, pgRouting, C# based on the ASP.NET framework.

*Geographical Information System*
Geographic Information System (GIS), captures, stores, analyzes, manages and presents data that is linked to location. Basically a GIS system handles geographical information in such a manner, as it integrates hardware, software, and data for capturing, managing, analyzing, and displaying all forms of geographically referenced information (Yusoff et al, 2008). In this project MapServer GIS software has been used to manage the geographical information such as roads and other features.

*Data Store*
For the Route Planning System to carry out its basic function, it needs a lot of data such as road segments, conditions of the road segments such as one-way, closed, congested etc., In addition to this operational data, it needs to store other administrative data such as user information, administrative data etc., PostgreSQL has been used in this projectfor as the backend database. PostgreSQL is a powerful, Open Source Object-Relational Database System (Blum, 2007). It is highly regarded for its reliability, data integrity, and correctness. It runs on all major operating systems, including Linux, UNIX (AIX, BSD, HP-UX, SGI IRIX, Solaris, Tru64), Mac OS X, and Windows.

*Route Planning*
The main objective of this project is to select the optimum driving route between two given points. In order to carry out this function, it needs to carry out several calculations and computations manipulating data with several perspectives. pgRouting provides functionality to PostGIS/ PostgreSQL (Patrushev, 2008). pgRouting has the capability to use Dijkstra, A* or Shooting Star in order to calculate the shortest path between any two given points. Dijkstra has the advantage of higher precision and fast response compared to other two methods and hence selected in this project (Patrushev, 2008; Schulzet al., 1999; Sniedovich, 2006; Stamp et al., 2002; Barati et al., 2008; Singh and swarup 2009).

*Development Platform*
ASP.NET, the next version of ASP (Active Server Pages), is a programming framework used to create enterprise-class Web Applications. The .NET framework consists of many class libraries and it can be used with different languages and has the capability of execution in different platforms. It is a very flexible framework and a developing Internet application with the .NET framework is very easy (Liberty et al., 2008). ASP.NET 3.5 was used in this project as the development platform as it provides better functionality implementation capabilities such as AJAX advanced features which can be used to develop user-friendly applications.

## User Management

The system developed supports two types of users, namely admin users and normal users, namely Administrators and Normal Users. Administrators have the privilege to input new data, modify data and other system parameters including information about road segments and delete data. Any ordinary user interested in accessing the system will come under the category of Normal Users. Anybody can register with the system and gain normal user privileges by providing his or her personal information such as username, password, full name, email address, phone number, residential address, closest city etc.,

In addition to the registered users, the system allows anonymous access too. Anonymous users can check driving directions between any two locations like registered users. But the result will be restricted to the current time of the day only. They will not be able to check the condition of the road network or the driving directions based on a future date or time.

The registered users can modify their personal information only. These users can change their information through the ASP.net form in the web module. The registered users can also perform a future search either for the current date and time or for a future date by giving a future date and time of travel and get the system to predict the best available shortest route for travelling on the particular date and time.

Once a user enters the required information, all the road segments between the locations entered are filtered out from the database. Then the Dijkstra's algorithm is applied through the pgRouting function library after finding out the nearest road segment's vertex points as source and target points. The algorithm uses an added reverse cost as a parameter in order to handle the roads being one-way or two-way. The one-way roads have a higher reverse cost value than two-way roads that is sufficient to support one-way streets without crashing the postgreSQL server. Then the complete path from the start location to the destination location is returned in a raster map image. This image is then delivered to the user via the web module as a web page or via the mobile module to the user's GPRS enabled mobile phone.

IV. ANALYSIS AND DESIGN

The system is made up of three main modules and two user sections. The modules are namely the Web Module, the Mobile Module and the Core Module. The main two user sections are Application User Section and the Admin User Section. Figure 6 shows the architecture of the system with inter-module communication.

Capturing and pre-processing of the user data is carried out by the user interface modules namely the Web Module and the Mobile Module, while the computation of the shortest path is handled by the Core Module. In addition, the Core module is responsible for the overall management of the system. This module connects with the MapServer and the database for retrieving the required data for computing the shortest path between two given locations. Backend database stores the road status data and user data. MapServer stores the map information.

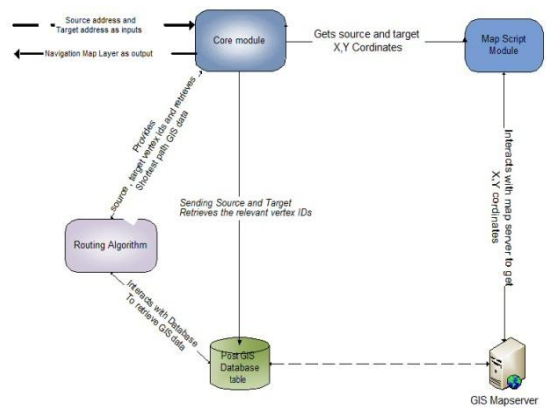

Figure 6: Architecture of the Route Advising System

Figure 7 shows the user interface in which a general web based user can access the system. Here the user has not logged onto the system, so he is not presented with the advanced search facilities. Figure 8 is the interface presented to the advanced user (logged in user), where the user can supply additional search parameters such as the date and time of intended travel.

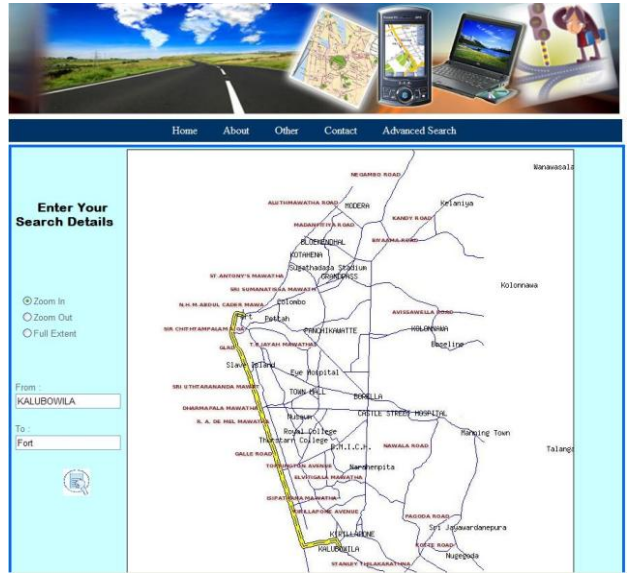

Figure 7: User Interface for Anonymous Users

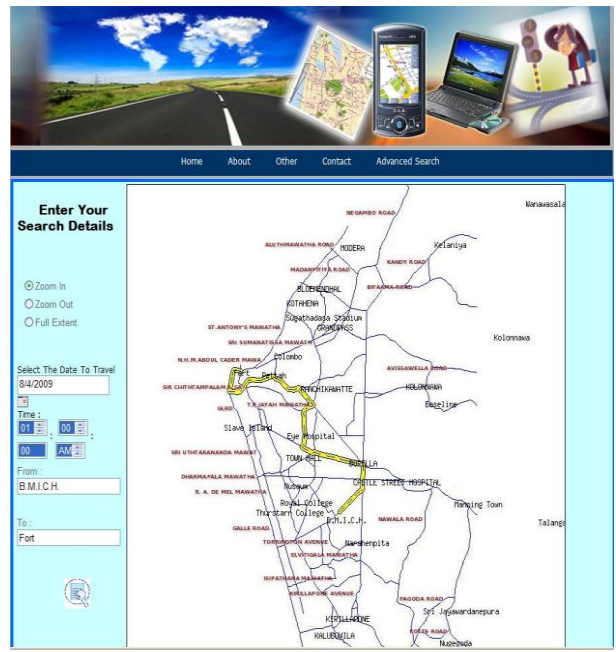

Figure 8: User Interface for Registered Users

Figure 9 shows the interface provided for a mobile user which can be accessed using a mobile phone with GPRS facility by entering the URL of the system.

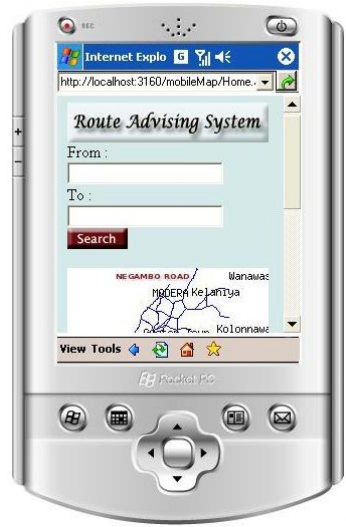

Figure 9: User Interface Provided for Mobile User

Figure 10 shows the administrator interface through which the system information can be modified. Administrator interface can be accessed only through the web as mobile module has the limited capability. Direct access to the database is prevented as the administration can be carried out only through the web interface provided. This ensures the security of the system against accidental changes the system.

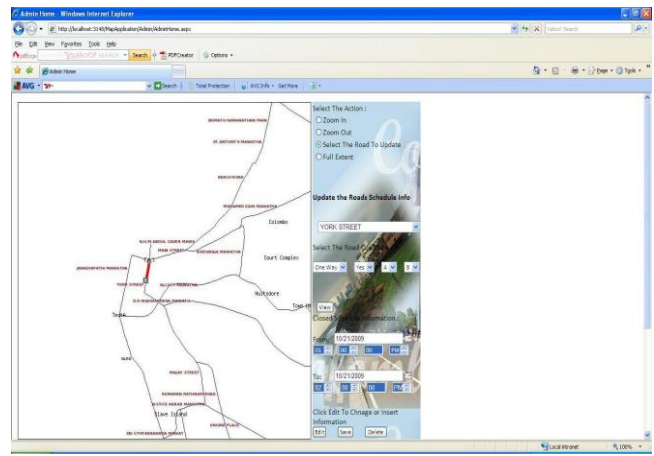

Figure 10: Administrator Interface on a Web Browser

## V. IMPLEMENTATION

This section describes the implementation details of the three main modules.

### Core Module

Core Module has been implemented integrating the MapServer, postgreSQL server, postGIS plugin, pgRouting along with Microsoft IIS server. the bin and lib files of the postgreSQL server as well. A sample vector map of the Colombo city has been used for testing. Figure 11 shows pseudo code used to implement the map file used for testing. This is the main file read by the map server to generate the output map in the projection with the mapping EPSG for Colombo. The generated map gets data from the vector files stored in the shape path of the map file.

```
Set NAME OF MAP
Set RESOLUTION
Set  EXTENT
Set UNITS
Load MAP
Load FONT
Load LIBRARY
WEB
   Load TEMPLATE
   Set IMAGE_PATH
   Set IMAGE_URL
END
Enable PROJECTION
END
```

Figure 11: Pseudo Code to Implement the Map

Figure 12 shows the pseudo code used to connect to the projectest database hosted on the postgrSQL server and select the roads layer in the map file. Different map feature is implemented on different layer of shape files. Using the shape file of main roads, the sql file is generated in order to create the table. The data in the table is then called through this layer object of map file. The names of the roads are marked using a LABELITEM in the file. Same method is used to display the shape data map information to the generated map image in web and the mobile modules.

```
LAYER
    Connect to the DBMS
    Send USERNAME & PASSWORD
Select DATABASE
    Select TABLE
    Load DATA
    Set STATUS
    Set TYPE
Mark (LABELITEM) ROADS
```

Figure 12: Code Segment to Select Layer in the Map File

Figure 13 shows the command used to convert the shape file into relations that can be stored in the relational database.

```
shp2pgsql -s 4015 -i -I "C:\towns.shp" towns > "C:\towns.sql"
```

Figure 13: Command used to Convert Shape Files to sql Files.

Figure 14 shows the pseudo code used to implement the Dijkstra's algorithm.

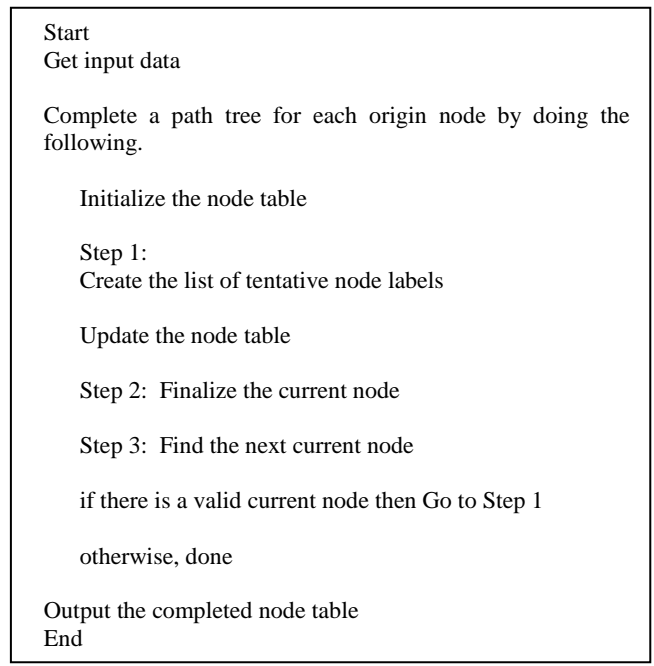

```
Start
Get input data

Complete a path tree for each origin node by doing the
following.

    Initialize the node table

    Step 1:
    Create the list of tentative node labels

    Update the node table

    Step 2:  Finalize the current node

    Step 3:  Find the next current node

    if there is a valid current node then Go to Step 1

    otherwise, done

Output the completed node table
End
```

Figure 14: Pseudo code for Implementing Dijkstra's Algorithm

The user input received from the user interface is converted into map coordinates for the purpose of further processing. This conversion is carried out using C# mapscripts running and implemented using mapscript.dll in the map server.

*Interface Modules*

User interfaces for the purpose of capturing user inputs and the presentation of the final results. Two different user interface modules were developed to serve on two types of devices, namely computers and mobile devices. The web module developed for the web based interface running on computers is a fully fledged user interface that support both normal users and administrators. Also, the output from this module consists of both a map and text based information giving the details of driving directions. In addition to, the output would also have additional features like zoom-in and zoom-out and customized search operations. The module was implemented using ASP.Net and C#.Net.

User interface for the mobile devices has been implemented on a module called the Mobile Module. This module can be accessed by mobile or wireless devices through General Packet Radio System or in short GPRS (Lempiäinen and Manninen, 2002). Since GPRS is supported by both 2G and 3G mobile networks, the application can be deployed even on an older mobile communication network. The mobile module has limited features compared to the web module due to inherent limitations of the web application supported in mobile devices. Figure 15 shows user activity flow chart in the Mobile Module.

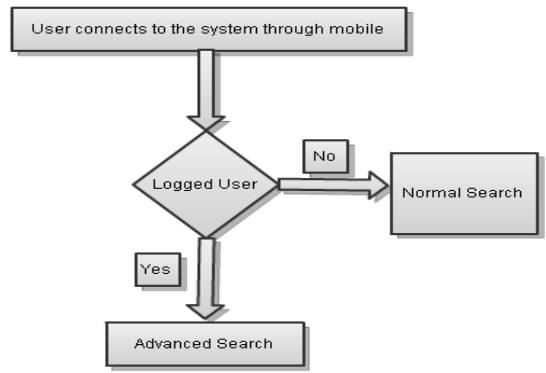

Figure 15: Mobile User Activity Flow Chart

## VI. EVALUATION

This section provides the evaluation of the prototype system developed. The prototype has been tested for usability in terms of ease of use and usefulness. A selected set of actual users who have familiar with web based applications were asked to test the features of the system. After the users interacted with the system for an extended period of time, they were asked to fill a detailed questionnaire based on their experience with the system. The questionnaire contained questions targeted towards each and every component (function) of the system. It also contained questions to capture details about the first experience, user friendliness, road and place information availability, correctness of the information, mobile module experience, user interface and overall performance.

As per Nielsen and Landauer (1993), 15 users would be sufficient for carrying out the evaluation of any software for usability. If more than 15 users were used, they would only detect the same issues that the others had already found and nothing new would be found. In this case, 20 users were used to carry out the test, in order to make sure all the issues were detected and documented properly. Figure 16 shows the personal profile of the test users.

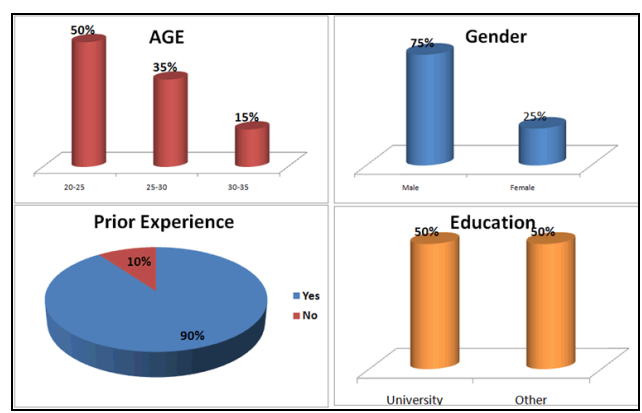

Figure 16: personal Profile of Test Users

From Figure 16, it can be seen that 90% of the users had prior experience in similar systems. Hence, the testers would automatically compare the features in the system with other systems that they had used before.

Reliability of system has been defined as the degree to which measures are free from errors and thus yield consistent results Zikmund (2003). A number of different reliability tests and coefficients have been used by researchers to ensure the reliability of the data collected. Split half reliability, Guttman, Parallel, Strictly Parallel and Cronbach's alpha are few of them. Cronbach's alpha is the most commonly used reliability coefficient as it can be interpreted as a correlation coefficient and ranges between 0 and 1 (Carmines and Zeller, 1979). In this study Cronbach's alpha is used to validate the reliability of each variable from the data collected. Table 2 shows the results of the reliability test carried out.

Table 2: Results of reliability Test

| Variables | No. of Items | Cronbach's Alpha |
|---|---|---|
| Ease of Use | 04 | .783 |
| Usefulness | 04 | .713 |

From Table 2 it can be seen that the Cronbach's alpha calculated for the variables Ease of Use and Usefulness are 0.783 and 0.713respectively. From this result, it can be concluded that the measures are all internally consistent and reliable as all of them have a Cronbach's alpha greater than 0.6.

Figure 17 shows the results of the survey. The Figure shows that majority of the users are either satisfied or very satisfied on all the aspects of the system. The results also show that 65% of the respondents are satisfied while 25% of the respondents are very satisfied.

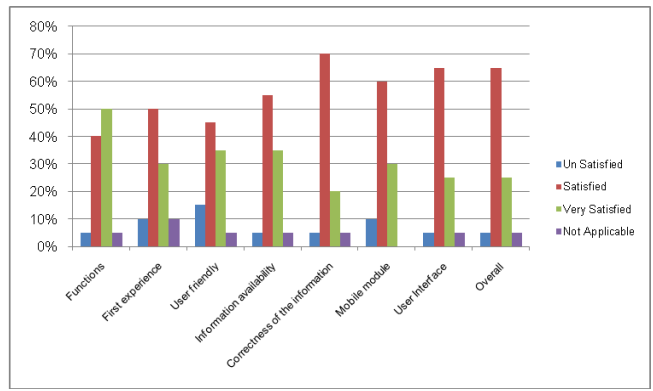

Figure 17: Results of the Evaluation for Different Functions

## VII. CONCLUSIONS AND FURTHER WORKS

This paper presents the development of a route planning system for the Colombo City with additional features such as planning on a future date and time and SMS and email alerts generated automatically in response to changes in the road network parameters.

The application has been implemented on both mobile and web platforms due to the high penetration of both technologies within the Sri Lankan population especially within main cities. To improve the user friendliness of the system while keeping cost of development at the lowest, open source technologies that provide the most advanced features were used. The technologies used in this project are namely Geographic Information System technologies, PostgreSQL database technologies, Mobile technologies, Web technologies, and Networking and Intelligent techniques such as Algorithms along with the additional techniques such as security techniques to secure the system.

The results of the survey carried out on the usefulness and the usability of the system proves that the main objectives of the project have been totally achieved.

This system should be extended to other major cities in Sri Lanka, finally covering the entire island. It is also proposed to enhance the system by including additional information like the public places like hospitals, schools, restaurants, cinemas etc.,

It is also proposed to include the time to travel between two locations as a parameter to enhance the system as what is more useful for the user would be the shortest time to travel between two locations rather than the physically shortest path between the locations. This would be very useful in cities like Colombo where traffic congestion is very high.